\documentstyle[prd,aps,preprint]{revtex}
\begin{document}
\draft

%
%  Uncomment following two lines and one below for 2 column format.
%@twocolumnfalse\endcsname

\preprint{KEK-TH-914}
\title{Color Ferromagnetism of Quark Matter 
and Quantum Hall States of Gluons in SU(3) Gauge Theory}
\author{Aiichi Iwazaki}
\address{Department of Physics, Nishogakusha University, Shonan Ohi Chiba
  277-8585,\ Japan.} 
\author{Osamu Morimatsu, Tetsuo Nishikawa}
\address{Institute of Particle and  Nuclear Studies, High Energy
  Accelerator Research Organization, 1-1, Ooho, Tsukuba, Ibaraki,
  305-0801, Japan}
\author{Munehisa Ohtani}
\address{The Institute of Physical and Chemical Research ( RIKEN ), 
2-1 Hirosawa, Wako, Saitama 351-0198, Japan}
\date{September 8, 2003} \maketitle
\begin{abstract}
\tightenlines
We show a possibility that a color ferromagnetic state
exists in SU(3) gauge theory of quark matter with two flavors.
Although the state involves three types of unstable
modes of gluons, all of these modes are stabilized by 
forming
a quantum Hall state of one of the modes. 
We also show that at large chemical potential, 
a color superconducting state ( 2SC ) appears even
in the ferromagnetic state. This is because
Meissner effect by condensed anti-triplet quark pairs 
does not work on 
the magnetic field
in the ferromagnetic state. 
\end{abstract}
%\pacs{73.61.-r,73.20.Dx,73.40.Hm,73.40.Gk}
%\hspace*{0.1cm}
\pacs{PACS 12.38.-t, 12.38.MN, 24.85.+p, 73.43.-f  \\ 
Quark Matter,\,\,Color Superconductivity,\,\, Quantum Hall States
\hspace*{0.5cm}}
%\vskip2pc
%%%%%%%%%%%%%%%%%%%%%%%%%%%%%%%%%%%%%%%%%%%%%%%%%
\tightenlines
One of the most intriguing phases possessed by
dense quark matter is the color superconductivity\cite{colors}
caused by the condensation of quark pairs with two flavors.
The condensation breaks the gauge symmetry of SU(3)
to SU(2). Consequently, some of
gluons gain masses due to the condensation
and Meissner effect operates in
the magnetic field of the gluons.
In the case of real superconductor,
the condensation of Cooper pairs of electrons
breaks U(1) gauge symmetry so that the magnetic field
associated with the U(1) gauge symmetry is 
expelled from or squeezed in the 
superconductor.
Therefore, both ferromagnetism and superconductivity
are not realized simultaneously in real 
condensed matter.
In the quark matter, however, both phenomena\cite{colorf}
can be realized simultaneously because
magnetic fields in remaining SU(2) gauge symmetry
are not affected by the condensation of the quark pair.

We have shown in the previous paper\cite{colorf} that
a color ferromagnetic state of quark matter 
arises in the SU(2) gauge theory, in which
a color magnetic field is generated spontaneously.
Although the magnetic field induces unstable modes of gluons\cite{savidy,nielsen},
the modes have been shown to be stabilized by the formation of
a quantum Hall state\cite{qh} of the unstable gluons. 
The quantum Hall state possesses a color charge.
The charge is supplied by quark matter.
Consequently, in the quark matter the stable ferromagnetic state in the SU(2) gauge theory exists
along with the quantum Hall state of the gluons. 

In this paper we point out in the SU(3) gauge theory that a stable color
ferromagnetic state\cite{savidy} of quark matter also 
exists along with a quantum Hall state of gluons.
In the state the diagonal color magnetic field
$\vec{B}\propto \cos\theta \,\,\lambda_3+\sin\theta \,\,\lambda_8$ is
spontaneously generated. 
The magnetic field induces three types of unstable modes, each of
which has different color charges. One of them condenses to form
a quantum Hall state. The condensation makes the ferromagnetic state
be stabilized. 
Furthermore, the state coexists  
with 2 flavor color superconducting state ( 2CS ). Namely,
In a ferromagnetic state with $B\propto \lambda_3$
( $\theta=0$ ),
an anti-triplet quark pair   
$\langle\epsilon^{3jk}q_jq_k\rangle$ can condense\cite{colors}. 
Since the magnetic field $B \propto \lambda_3$ does not couple with the quark condensation,
the Meissner effect does not operate on the field.
In this way
the coexistence of both ferromagnetism and superconductivity 
is possible in the SU(3) gauge theory. This is owing to the fact that
the SU(3) gauge group has
the maximal Abelian subgroup of U(1)$\times$U(1);
each group of U(1) is associated with superconductivity or 
ferromagnetism, respectively.  
There is a critical chemical potential which separates the phase of 
only the ferromagnetic state without the superconductivity 
and the coexistence phase 
of both states. 

We assume in the paper that loop calculations
or perturbative calculations give physically correct results
in the ferromagnetic phase 
although the gauge coupling constant is not necessarily small
in the energy regime ( $500\,\,\mbox{MeV}\sim 1000$ MeV ) of our consideration.

As is well known\cite{savidy,nielsen}, 
the one loop effective potential $V$ 
of the constant color magnetic 
field in the SU($n_{\rm c}$) gauge theory with $n_{\rm f}$ flavors
is given by
$V_{\rm eff}= \frac{11N}{96\pi^2}g^2B^2\left(\log
  (gB/\Lambda^2)-\frac{1}{2}\right)-\frac{i}{8\pi}g^2B^2$,
with an appropriate renormalization\cite{nn} of the gauge coupling $g$,
where $N=n_{\rm c}-2n_{\rm f}/11$. 
Hereafter we consider the SU(3) gauge theory with 
massless quarks of two flavors; $N=29/11$. But,
most of our results ( e.g. the existence of 
the ferromagnetic phase ) hold even for massive quarks.
The potential implies the spontaneous generation of
a color magnetic field. The direction of the magnetic field in real space 
can be arbitrarily chosen and the direction in color space can be taken in general such as
$B$ is in the maximal Abelian sub-algebra; $B=|B|(\cos{\theta}\,\,\lambda_3+\sin{\theta}\,\,\lambda_8)$,
where $\lambda_i$ are Gell-Mann matrices and we restrict the value
of $\theta$ such as $-\pi/6\leq \theta \leq\pi/6$ due to the Weyl symmetry.
In any case of their choices the spontaneous generation of
the magnetic field breaks the spatial rotational symmetry and the
SU(3) gauge symmetry into  
the gauge symmetry of U(1)$\times$U(1). 
It is interesting to note that the hypothesis of ``Abelian dominance''\cite{abelian} 
holds exactly in this ferromagnetic phase since 
physics at long wave length is governed by only the gauge fields
of maximal Abelian gauge group U(1)$\times $ U(1). 

Although 
the magnetic field is spontaneously generated, this state 
is known to be 
unstable due to
the presence of the imaginary part in $V_{\rm eff}$. Namely,
there are unstable modes\cite{nielsen,nn} of gluons in the ferromagnetic state.
These modes generate the imaginary part in $V_{\rm eff}$.
In general, unstable modes are excited to lead to  
a stable state by their condensation.
It is, however, non-trivial
to find the stable state of gluons in the gauge theory.   

As we have shown in the SU(2)
gauge theory\cite{colorf}, such
unstable gluons occupy the lowest Landau level 
and condense to form
a fractional quantum Hall state. Quantum Hall states 
of electrons\cite{qh} are known
to have a finite gap like BCS states and be stable. Similarly, the 
quantum Hall state of the gluons has been shown to be 
stable; 
unstable modes disappear in the quantum Hall state. 
(According to numerical simulations\cite{naka},
Laughlin states representing fractional 
quantum Hall states
are shown to arise even
for the bosons just like gluons.)

The condensed gluons leading to the quantum Hall state
possess a color charge, in other words, the color charge
condenses in the state.
Such a color charge of the condensed gluons must be supplied by
others in a neutral system. 
Quark matter is such a supplier.
Therefore, 
the stable ferromagnetic state involving the quantum Hall state of gluons
can arise in the dense quark matter. (Since the energy density of the quarks
in the presence of the magnetic field is lower\cite{colorf} than that of the quarks 
without the magnetic field, the spontaneous generation of the magnetic field
is favored also in the quark sector.)
In contrast,
the state does not arise as a spatially 
uniform vacuum state of the gauge 
theory. There are no suppliers of the color charges
in the vacuum.
The unstable modes cannot form quantum Hall states. Instead,
their large fluctuations would lead to
``spaghetti vacuum''\cite{nn,spa} in which
quark confinement would be realized.

Now, we wish to study unstable modes of gluons in the SU(3) gauge theory
and to see how they are stabilized by forming a quantum Hall state.
In the SU(3) gauge theory there are
three types of unstable modes in general. 
They are identified as modes 
having an imaginary part in their energy spectra. In general, 
spectra of the gluons in the magnetic field $B$ are given by $E^2=2g_iB(n+1/2\pm 1)+k_3^2$
where $g_i$ is the coupling strength with the magnetic field.
$n$, $\pm 1$ and $k_3$ denote Landau level, spin and momentum parallel
to $\vec{B}$,
respectively.
The imaginary part in $E$ comes from the contribution of 
their anomalous magnetic moments,
which corresponds to the negative term in $E^2$. 
Obviously, the unstable modes are those occupying the lowest Landau
level ( $n=0$ ) with spin parallel ( $-1$ ) to $\vec{B}$
and with small momentum $k_3^2<2g_iB$ .
Hence, the identification of the unstable modes can be
done by looking for such modes with the anomalous magnetic moments.
In general, there are $6$ gluons $A'_{\mu}$ which can 
couple with $B$, i.e. $[A'_{\mu},B]\neq 0$. Thus, $3$ complex fields
can be composed of the $6$ real fields of the gluon. They are color charged
vector fields coupled with $B$ and 
possible unstable modes. 
Actually, unstable modes are easily identified from the SU(3) gauge
fields as,

\begin{equation}
\Phi_1=(A_1+iA_2)/\sqrt{2},\,\,
\Phi_2=(A_4+iA_5)/\sqrt{2},\,\,\mbox{and}\,\,
\Phi_3=(A_6-iA_7)/\sqrt{2},
\end{equation}
where $A_a$ is defined as a spatial component of the gauge field with particular polarization;
$A_a^{\mu}=e^{\mu}A_a$ with $e^{\mu}=(0,1,i,0)$.
These modes occupy the lowest Landau level with spin
parallel to the magnetic field. 
They have conserved charges of U(1)$\times$U(1) such that
$(Q_3(\Phi_1)=1,\,Q_8(\Phi_1)=0)$,
$(Q_3(\Phi_2)=1/2,\,Q_8(\Phi_2)=\sqrt{3}/2)$
and $(Q_3(\Phi_3)=1/2,\,Q_8(\Phi_3)=-\sqrt{3}/2)$.
To clarify that these modes
are really unstable, we
see a Lagrangian for these modes by extracting them from 
the action of the SU(3) gauge field; 
we simply neglect the other stable modes,

\begin{equation}
\label{L}
L_{\rm unstable}=\sum_{s=1,2,3}(|(i\partial_{\mu}-g_sA_{\mu}^{B})\Phi_s|^2+2g_sB|\Phi_s|^2)-V(\Phi)
\end{equation}          
where $g_1=g\cos\theta$, $g_2=g(\cos\theta+\sqrt{3}\sin\theta)/2$,
$g_3=g(\cos\theta-\sqrt{3}\sin\theta)/2$ 
 and the potential,
$V(\Phi)=g^2((\sum_{s=1}^3|\Phi_s|^2)^2-3|\Phi_2\Phi_3|^2)$.
$A_{\mu}^{B}$ represents a gauge potential of the magnetic field $B$.
We have used the fact that the modes $\Phi_s$ occupy the lowest Landau level.
We should note that all of $g_i$ are positive or zero in the range of 
$-\pi/6\leq\theta\leq\pi/6$.

We can see the negative mass terms ( $-2g_sB|\Phi_s|^2$ ) 
of the modes,
$\Phi_s$. It implies the instability of the state, $\langle\Phi_s\rangle=0$.
Thus, they are unstable modes.  
This situation is very similar to a model of complex scalar field
with double well potential. Namely, in the model with a 
potential such as $-m^2|\Phi|^2+\lambda |\Phi|^4/2$, the state such that
$\langle\Phi\rangle=0$ is unstable. Thus, 
a spatially uniform unstable mode ( the mode is given by the fluctuation,
$\delta\Phi$ with zero momentum, $\vec{k}=0$, 
 around the state $\langle\Phi\rangle=0$ ) is excited to condense and form
a stable state $\langle\Phi\rangle=m/\sqrt{\lambda}$. 
In this simple model the spatially uniform unstable
mode $\delta\Phi(\vec{k}=0)$ is unique. 
On the other hand, there is no such a solution
as $\langle\Phi_s\rangle=$ constant $\neq 0$ in the gauge theory described by eq(\ref{L}). 
This is because the gauge potential $A_{\mu}^B$ has a spatial dependence. 
Physically, not only there are no spatially uniform modes
but also 
there are infinite number of 
degenerate unstable modes in the magnetic field.
That is, the unstable modes $\Phi_s$ in the lowest
Landau level have such a form of the wave functions
as $\exp[-ik_2x_2-\frac{1}{2}\,g_sB(x_1-k_2/g_sB)^2]$,
where $k_2$ denotes a momentum perpendicular to $\vec{B}=(0,0,B)$.
Here we have taken only modes with $k_3=0$ uniform in the
direction parallel to $B$. Obviously, all of 
the modes labeled by the parameter $k_2$ 
are infinitely degenerate. The parameter indicates the location
of the modes in the coordinate of $x_1$.  We can see that 
none of these unstable modes is spatially uniform.
Since each type of the
modes involves infinitely degenerate non-uniform states, it is not trivial to find
a uniform stable state formed by the condensation of the modes.   
Actually, it was a very difficult task to find such a stable
state formed by the unstable modes. We need to take into account 
repulsive interactions $V(\Phi)$ between the modes in order to
find the state. Although a reasonable variational state has been
proposed\cite{nn}, the state has no spatial uniformness and has not yet been
shown to be really stable. 

In the case of the SU(2) gauge theory, only one type of the unstable
mode exist with a repulsive interaction term such as $|\Phi|^4$. 
We have shown that the mode forms a quantum Hall state,
in which the instability disappears.
On the other hand, in the SU(3) gauge theory 
we have three types of the unstable modes
with their repulsive interaction terms $V(\Phi_s)$ 
more complicated than the term $|\Phi|^4$ in the case of
the SU(2) gauge theory. But we can show that
one ( $\Phi_1$ ) of the modes 
condenses to form a stable quantum Hall state. 
Consequently the instability of the mode disappears just as 
in the SU(2) gauge theory.
The other modes ( $\Phi_{2,3}$ ) 
gain sufficiently 
large positive mass terms from
the condensation of $\Phi_1$. Since such positive masses are larger than 
the bare negative masses, the modes obtain positive mass terms in 
consequence.
In this way all of the unstable modes are stabilized. 

It is important to 
note that the relevant unstable modes should be uniform 
in a coordinate of $x_3$. Namely, the mode with
$k_3=0$ has stronger instability than those with $k_3\neq 0$
since it grows up more rapidly in time;
the wave function of the mode with
$E(k_3)=i\sqrt{2g_sB-k_3^2}$ grows up just as $e^{|E(k_3)|t}$. 
Thus, we take only such a mode with $k_3=0$.
Accordingly, the gluons of
the field $\Phi_s$ are spatially 2 dimensional ones just like
electrons in 2 dimensional quantum wells, which may
form quantum Hall states under external magnetic field.
Consequently, we can use a Chern-Simons gauge theory in spatial 
2 dimensions for the discussion of the unstable modes 

It is well known that the Chern-Simons gauge theory\cite{qh}
is very useful 
for analysis of quantum Hall states. That is, we consider
so called composite gluons which are bosons with
fictitious magnetic flux. The flux is expressed by 
the Chern-Simons gauge field. The point for the use of such composite
gluons is that only when the fictitious flux cancels with the magnetic field
$B$, the bosons can condense to form quantum Hall states
with spatially uniformness. This way of understanding the quantum Hall
state is well known in condensed matter physics;
the picture of ``composite electrons'' is used as very powerful 
tool for the analysis of quantum Hall states. 
Thus, we 
apply the method to the physics of the gluons.
        
Using Chern-Simons gauge fields, we write down spatially 2 dimensional
Lagrangian of the composite bosons representing the unstable modes in the
following way, 

\begin{eqnarray}
L_a=|(i\partial_{\mu}-g_1A_{\mu}^B+a_{\mu})\phi_1|^2+2g_1B|\phi_1|^2
-\frac{\epsilon^{\mu\nu\lambda}}{4\alpha}a_{\mu}\partial_{\nu}a_{\lambda} 
\nonumber\\
+\sum_{s=2,3} (|(i\partial_{\mu}-g_sA_{\mu}^B)\phi_s|^2+2g_sB|\phi_s|^2)-
V_a(\phi)
\label{eq} 
\end{eqnarray}
with $V_a=V/l_3$ ( $l_3$ being the length scale 
of quark matter in 3 direction ) and $\phi_s=
\Phi_s \sqrt{l_3/2}$ for $s=2,3$.
We have introduced a Chern-Simons gauge field only for
the unstable mode, $\phi_1$, which has the largest negative mass term
and is expected to compose quantum Hall states.
This Lagrangian describes the composite bosons $\phi_1$ attached by
Chern-Simons flux $\epsilon^{ij}\partial_ia_j$ ( fictitious magnetic
flux ).
$\alpha$ should be chosen to be 2$\pi\times $integer
in order to guarantee that the field $\phi_1$ describes boson. 
Hereafter, we take
$\alpha=2\pi$ for simplicity.
The equivalence of this Lagrangian, $L_a$ and $L_{\rm unstable}$
has been demonstrated in the operator formalism\cite{seme} although 
the equivalence had been known in the path integral formalism using the world
lines of the $\phi_a$ particles.

In order to see that the unstable modes denoted by $\phi_1$ form
a stable quantum Hall state, we derive equations of motion for the
field $\phi_1$,

\begin{eqnarray}
\label{eq1}
\phi_1^{\dagger}\,i\partial_0\,\phi_1+{\rm c.\,c.}+2a_0\,|\phi_1|^2=\frac{1}{4\pi}\epsilon_{ij}\,\partial_i\,a_j
\\
\phi_1^{\dagger}\,(\,i\partial_i-g_1A_i^B+a_i)\,\phi_1+{\rm c.\,c.}=\frac{1}{4\pi}\epsilon_{ij}(\partial_0\,a_j-\partial_j\,a_0) 
\\
(i\,\partial_0+a_0)^2\,\phi_1-(\,i\vec{\partial}-g_1\vec{A}^B+\vec{a}_1\,)^2\,\phi_1+2g_1B\,\phi_1=\frac{\partial}{\partial\phi_1^{\dagger}}V_a.
\end{eqnarray}    

We can easily find a solution of a spatially uniform state $\langle\phi_1\rangle=v_1$ 
( $\langle\phi_2\rangle=\langle\phi_3\rangle=0$ ) only when the Chern-Simons flux 
cancels with $B$, that is, $g_1\vec{A}^B=\vec{a}$. This
cancellation is only possible for $a_0$ and $v_1$ satisfying

\begin{equation}
\label{eq2}
2a_0v_1^2=\frac{g_1B}{4\pi} \quad \mbox{and} \quad  a_0^2v_1+2g_1Bv_1=
\frac{\partial V_a}{\partial\phi_1^{\dagger}}=\frac{2g^2}{l_3}v_1^3 
\end{equation}
or in dimensionless notations
\begin{equation}
2\bar{a}_0\bar{v}_1^2=\frac{\cos\theta}{4\pi} \quad \mbox{and} \quad
\bar{a}_0^2+2\cos\theta=\frac{2g^2}{\bar{l_3}}\bar{v}_1^2
\label{eq3}
\end{equation}
where $\bar{a}_0(\theta),\,\,\bar{v}_1^2(\theta)$ and $1/\bar{l}_3$ are dimensionless
quantities normalized in unit of $(gB)^{1/2}$. 

Solving the equations (\ref{eq2}) or (\ref{eq3}) for $v_1$ and $a_0$, and inserting $v_1$ into the 
potential $V(\phi)$, 
we can see that the modes $\phi_2$ and
$\phi_3$ gain a sufficiently large positive mass ( $=2g^2v_1^2/l_3=2g_1B+a_0^2$ )
caused by
the condensation of the mode $\phi_1$. Therefore, the masses of $\phi_{2,\,3}$ become positive
owing to this Higgs mechanism because $2g_1B+a_0^2$ is larger than 
their original negative masses, $-2g_2B$ and $-2g_3B$,
respectively; note that $g_1(\theta) >g_2(\theta),\,g_3(\theta)$ for $-\pi/6<\theta <\pi/6$. 
Hence, 
the instability of the state, $\langle\phi_{2,\,3}\rangle=0$
is removed. As in the SU(2) gauge theory, the state $\langle\phi_1\rangle=v_1$
is also stable. In this way, the condensation of the field, $\phi_1$
leads to the stable ferromagnetic state.

It is well known in the picture of 
the composite boson that the state $\langle\phi_1\rangle=v_1$
represents
a quantum Hall state of the $\phi_1$ particles 
with so called filling factor, $\nu=2\pi\rho_{g1}/g_1B$, being equal to $1/2$;
$\rho_{g1}$
( $=\phi_1^{\dagger}\,i\partial_0\,\phi_1+{\rm c.\,c.}+2a_0\,|\phi_1|^2$ ) denotes a number density of  $\phi_1$ particles.
It is easy to show\cite{zhang} that Hall conductivity of the state is given by $2\pi\,\nu/g_1 $
as expected in usual quantum Hall states. Similarly, we can show that the state has a finite gap,
$E$, by solving a small fluctuation around $\phi_1=v_1$ in eq(\ref{eq1}); $E=2\sqrt{a_0^2+g_1^2v_1^2/l_3}$.
Therefore, all of the unstable modes become harmless owing to
the formation of a quantum Hall state of the mode $\phi_1$.
Obviously, the result holds irrespective of quark mass.

Next, we wish to determine the direction $\theta$ 
of $B$ in the color space. The determination is not trivial. 
In order to determine $\theta$ explicitly,
we need to 
take account of the physical conditions, that is,
color neutrality of quark matter
and the minimum of its energy including the energy of the gluon condensate; 
both the energies of the quark and 
the condensate, $\langle\phi_1\rangle$, of the unstable modes depend on the direction.

We first impose the color neutrality conditions,
$\langle\lambda_3\rangle=\langle\lambda_8\rangle=0$, where 
the average should be taken 
over all of the quarks and the gluon condensate;
we should note that owing to the generation 
of $B$, the SU(3) gauge symmetry is broken into 
the gauge symmetry of U(1)$\times$U(1) so that only conserved color
charges are those associated with the group
generators of $\lambda_{3,\,8}$.
It is easy to see that
the condition implies that 
number density $\rho_i$ of quarks with color type $i$
satisfies the following equations, $\rho_1+\rho_{g1}=\rho_2$ and
$\rho_1+\rho_2=\rho_3$.
(We denote the color types of quarks such that $q_1=(1,0,0),\,\,q_2=(0,1,0)$ and
$q_3=(0,0,1)$.)

Furthermore, we impose the condition of the energy minimum. 
In order to calculate the energy of quarks, we note that the coupling strength
$e_i$ of $i$-th quark with the magnetic field is given by $e_1=g(\cos{\theta}+\sin{\theta}/\sqrt{3})/2$,
$e_2=g|-\cos{\theta}+\sin{\theta}/\sqrt{3}|/2$, and $e_3=g|-2\sin{\theta}/\sqrt{3}|/2$,
where $g$ is the gauge coupling constant.
Then, the energy of $i$-th massless quark in the magnetic field is given by 
$\sqrt{e_iB(2n+1\pm 1)+k_3^2}$ where integer $n\geq 0$ denotes Landau level and
$\pm 1$ does the quark spin; momentum $k_3$ is a component parallel to $B$.  
Minimizing the total energy density of the three types of quarks in
$\theta $ is 
difficult in general. As a simple example, we calculate it in a case where
all of quarks occupy the Lowest Landau level ( $n=0$ ) with spin
parallel ( $-1$ ) and the condensation of gluons $\rho_{g1}$ is negligible.
Then, the energy density of $i$-th quark is given by
$\epsilon_i=n_{\rm f}e_iBk_{{\rm f}i}^2/4\pi^2=\pi^2\rho_i^2/n_{\rm f}e_iB$ where
$k_{{\rm f}i}$ denotes
the Fermi momentum and $n_{\rm f}=2$ denotes the number of flavors. Here we have used the expression of 
the number density of quarks, $\rho_i$,
in terms of the Fermi momentum 
$k_{{\rm f}i}$; $\rho_i=n_{\rm f}e_iBk_{{\rm f}i}/2\pi^2$. Since 
all of the quark densities are identical, 
$\rho_i=\rho$, when $\rho_{g1}=0$, 
we find that the total energy density, 
$\sum_{i=1\sim 3}\epsilon_i=\pi^2\rho^2\,(1/e_1+1/e_2+1/e_3)/n_{\rm f}B$,
takes the minimum value at $\theta=\pi/6$. This is a simple case
of all quarks occupying the lowest Landau level ( namely much small $\rho/(gB)^{3/2}\ll1$ )
and of the negligible gluon contribution. 
But actually, the gluon contribution to the energy is dominant
over the quark contribution since
the energy density of the quarks 
occupying higher Landau levels depends very weakly on
the direction $\theta$ of the color magnetic field.

In order to see it
we calculate the energy of the gluon condensate depending on 
$\theta$. Using the solutions of eq(\ref{eq3}), we find that 
the energy density of the condensate normalized by $(gB)^{2}$ is given by

\begin{equation}
\bar{E}_v=(\bar{a}_0^2(\theta)\bar{v}_1^2(\theta)-2\bar{v_1}^2(\theta)\cos\theta+
\frac{g^2}{\bar{l}_3}\bar{v}_1^4(\theta))/\bar{l}_3=
(2\bar{a}_0^2(\theta)\bar{v}_1^2(\theta)-\frac{g^2}{\bar{l}_3}\bar{v}_1^4(\theta))/\bar{l}_3,
\end{equation} 
where we have represented the dependence of $\theta$ explicitly. Note that 
we have the factor of $1/\bar{l}_3$ in the above equation in order to obtain the three demensional energy
density, $\bar{E}_v$, from two dimensional lagrangian in eq(\ref{eq}).
We have proved numerically that the energy of the condensate takes the minimum  
at $\theta=0$.
Actual values of $\bar{E}_v$ 
at $\theta=0$ are as follows.
$\bar{E}_v\bar{l}_3=-0.02, \quad -0.08,\quad -1, \quad -10 \quad \mbox{for} 
\quad g^2/\bar{l}_3\,(\,=g^2/(l_3\sqrt{gB})\,)\,=20,\quad 10,\quad 1, \quad
0.1$, respectively. Smaller coupling constants, $g^2/\bar{l}_3$, give rise to 
negatively larger energies.   We do not know how large the coupling parameter
$g^2$ is. We tentatively assume that $g^2$ is of the order of $1$ or less
for the consistency of our calculation.
Then, assuming $l_3=1\sim 3\,\, \mbox{fm}$ in real
quark matter produced by heavy ion collisions and assuming $\sqrt{gB}\sim
200\mbox{MeV}$, we obtain $\bar{l}_3=1\sim 3$ and the order of the magnitude of the coupling constant
$g^2/\bar{l}_3$ being $O(0.1)\sim O(1)$.
Hence, the energy of the condensate takes a value, at least, such as $\bar{E}_v\simeq-1$
at $\theta=0$.
On the other hand,
the energy density, $E_{B}$, of the quarks normalized by $(gB)^2$ is approximately
given by $E_{B=0}=(3\pi^{2/3}/2^{7/3})(\rho/(gB)^{3/2})^{4/3}\simeq
1.28\,(\rho/(gB)^{3/2})^{4/3}$, $i.e.$ the energy density of the quarks in the absence of the magnetic
field. Namely, $E_{B}$ is nearly equal to $E_{B=0}$ for large quark
number density $\rho$; $E_{B}=E_{B=0}$ in the limit of
$\rho/(gB)^{3/2}\to \infty$. Note that $E_{B=0}$ does not depend on
$\theta$. This indicates that $\theta$ dependent part of $E_{B}$
is much small. For instance, $E_{B}-E_{B=0}$ at $\theta=0 $ is about $0.09$ ( $0.13$ ) for  
$\rho/(gB)^{3/2}=10$ ( $1000$ ).
Therefore, $\theta$ dependence of the energy is determined by 
$\bar{E}_v$. Thus, we conclude that
the direction of the magnetic field $B$ is chosen 
such as $\theta=0$, which gives the minimum energy of 
the quarks and the gluons.
Anyway, the condensation of $\phi_1$ makes $B$ point to the direction of 
$\lambda_3$ in maximal Abelian subalgebra.

A comment is in order. To make a quantum Hall state, we can
make all of the fields $\phi_s$ condense by introducing different Chern-Simons
gauge fields, $\vec{a}_s$ to each $\phi_s$. Then, we have such a state 
as $\langle\phi_s\rangle=v_s\neq 0$ for $s=1\sim 3$. But we can show that the state is
not stable because the fluctuations $\delta\phi_s$ have an imaginary
frequency in their spectra. On the other hand, the formation of the quantum Hall state
of only the field $\phi_1$ makes the state 
$\langle\phi_1\rangle\neq 0,\,\,\langle\phi_2\rangle=0$
and $\langle\phi_3\rangle=0$ stable as we have explained above. 
Similarly we can show that a quantum Hall state such as $\langle\phi_1\rangle=0$, $\langle\phi_2\rangle\neq
0$ and $\langle\phi_3\rangle\neq 0$, is also stable. In the state two fields
$\phi_{2,3}$ condense. Thus, the state possibly exists although
we consider only the case of the condensed state of $\phi_1$ 
in this paper.

Up to now, we have shown that there are three types of the unstable
modes of gluons
induced by the color magnetic field generated spontaneously. 
These unstable
modes are stabilized by the formation of the quantum Hall state of the
unstable mode $\phi_1$. Owing to the generation of the magnetic field,
the gauge symmetry SU(3) is broken
into the gauge symmetry U(1)$\times$U(1). Furthermore,
the formation of the quantum
Hall state ( $\langle\phi_1\rangle=v_1$ ) breaks the gauge symmetry into the gauge
symmetry of U(1). This is the case for the color magnetic 
field whose direction, $\theta$, is
general. On the other hand, one of the unstable modes disappears
at specific values of $\theta$. For example, $g_3$ vanishes
for $\theta=\pi/6$ so that the negative mass ( $-2g_3B$ ) of the field $\phi_3$
vanishes. $\phi_3$ becomes massless. Thus, the mode of $\phi_3$
is stable.  
In this case, the gauge symmetry SU(3) is broken into the
gauge symmetry SU(2)$\times$ U(1). Namely, there are 4 massless real fields;
gauge fields of the maximal Abelian gauge group and $A_{\mu}^{6,7}$
corresponding to $\phi_3$.
Subsequent formation of a quantum Hall state with $\langle\phi_1\rangle=v_1$
breaks the symmetry into the gauge symmetry of U(1). Stability of
the state $\langle\phi_{2,\,3}\rangle=0$ is guaranteed by the positivity of their
masses produced by the condensation of the field $\phi_1$;
$\phi_3$ becomes massive as well as $\phi_2$.
In any case, stable ferromagnetic states accompanied by quantum Hall
states are realized in the SU(3) gauge theory. Since the gauge
symmetry is broken, the state is not a real vacuum of QCD,
but, a state of quark  matter.

We should note that the color charge $Q_3(\phi_1)$ of the 
condensate $\langle\phi_1\rangle$ must be supplied by the quark matter
produced by heavy ion collisions.
Unless the chemical potential of the quark matter is large enough to supply the charge,
such a condensation cannot occur, so that the stable ferromagnetic state
cannot arise. Thus,
we wish to estimate the minimum chemical potential of the quark matter
 needed for the formation of the condensate.
We note that two-dimensional color charge density of the gluon condensation is given by $\rho_{g1}=g_1B/4\pi$, namely
the number density of the $\phi_1$ particles. When the radius of the quark matter is $l_3$,
the three dimensional density, $\rho_g$, of the condensed gluons is 
approximately given such that $\rho_g=\rho_{g1}/l_3$. On the other hand, 
the color charge density of u and d quarks is $\frac{1}{2}\times
2\times k_{\rm f}^3/3\pi^2$ where
$k_{\rm f}$ denotes a Fermi momentum; the factors $1/2$ and $2$ 
come from the coupling strength in unit of $g$ and the number of the
flavor, respectively. The chemical
potential, $\mu$, of the quark at zero temperature is given by $\sqrt{k_{\rm
    f}^2+m_q^2}$ where we have taken account of the quark mass,
$m_q\sim 300$MeV. 
Hence, by equating both densities we obtain the lower bound of the chemical potential such that 

%\begin{eqnarray}
%&\mu&=\sqrt{((3\pi g_1B/4l_3)^{1/3})^2+m_q^2}, \\
%&\sim& 350\mbox{MeV}\sqrt{(180/350)^2(\langle gB \rangle /0.04\,\mbox{GeV}^2)^{2/3}(3\,\mbox{fm}/l_3)^{2/3}+(300/350)^2(m_q/300\mbox{MeV})^2}, 
%\end{eqnarray}
\begin{eqnarray}
\mu &=&
\sqrt{\left(\left({3\pi g_1B}/{4l_3}\right)^{1/3}\right)^2+m_q^2} 
\nonumber \\ &\sim& 350 \,
\mbox{MeV}\sqrt{\left(\frac{180}{350}\right)^2\frac{(\langle gB \rangle
/0.04\,\mbox{GeV}^2)^{2/3}}{(l_3/3\,\mbox{fm})^{2/3}}
+\left(\frac{300}{350}\right)^2(m_q/300\mbox{MeV})^2} 
\end{eqnarray}
where we have referred to a typical scale of QCD as $\langle gB \rangle$.
Therefore, the ferromagnetic state of the quark matter may arise 
in heavy ion collisions
at chemical potentials
larger than $350 \,
\mbox{MeV}$.

We have discussed the ferromagnetic phase of the
SU(3) gauge theory.
The phase appears at chemical potentials larger than a critical one, 
below which the hadronic phase is
present. 
We already know the presence of
a color superconducting phase ( 2SC ) in much large chemical potential.
Thus, we wish to ask which state arises in the quark matter,
the color superconducting state or
the color ferromagnetic state. 
Although both states do not appear simultaneously in 
the ordinary matter owing to Meissner effects, 
they are compatible in the SU(3) gauge theory
without any contradictions.

In order to analyze the possibility, we note that the direction of the color magnetic field
is pointed such as $\theta=0$. Thus,
an anti-triplet quark pair condensation such as 
$\epsilon^{ijk}\langle q_jq_k\rangle=(0,0,u_3)$ 
may arise because the magnetic field $B\propto \lambda_3$
is not inhibited by the Meissner effect;
the magnetic field does not couple with the condensation.
Therefore, both states may arise simultaneously 
in a dense quark matter.

Explicit calculations have been done\cite{ebert} within a NJL model
where the effects of the color magnetic field, $B\propto \lambda_3$
on the chiral condensation and the quark pair condensation
have been discussed. The authors of the paper
considered such a model in order to see
the effect of vacuum
fluctuations of the color magnetic field, $\langle B^2\rangle\neq 0$ on the quark matter.
On the other hand, as we have shown, such a color magnetic field is generated spontaneously.
Using their results, we find that 
the magnetic field induces the chiral condensation at any chemical potentials less than a critical 
one, beyond which the quark pair condensation arises; 
2CS appears.
It is interesting that the chiral condensation 
is not compatible with the quark pair condensation. Thus,
in the 2CS, the chiral
symmetry is not broken. The point we learn from the paper is
that the ferromagnetic phase can coexist with the color
superconducting phase.

It seems apparently that the pointing of $B$ to $\lambda_3$ is
necessary for the existence of the superconducting state. 
We remember that our conclusion of $\theta=0$ has been obtained by
minimizing the energy of the quark matter  
under the assumption
of $g^2$ being the order of $1$. Without the assumption the conclusion
of $\theta=0$ cannot be obtained. However, the magnetic field
naturally orients to $\lambda_3$ when the superconducting state
arises in the ferromagnetic state at large chemical potentials. This is 
because the presence of such a state is energetically 
more favored than the absence of the state; the state with $\theta=0$
involving the superconducting state is energetically favored than a
state with $\theta\neq 0$ involving no superconducting state.

To summarize, we have found that the phase of the quark matter at zero
temperature has the following structure. At small chemical potentials
the hadronic phase with the broken chiral symmetry is present.
It is unclear whether or not the ferromagnetic phase arises before
the chiral symmetry is restored when we increase the chemical
potential. But it is clear that the hadronic phase is changed into
the ferromagnetic phase at a certain chemical potential.
In the phase the color magnetic field, 
$\vec{B}\propto \cos\theta \,\,\lambda_3+\sin\theta \,\,\lambda_8$,
is generated spontaneously
whose direction in the color space is pointed such as $\theta=0$.
A quantum Hall state of gluons, $\phi_1$, is also formed
in the phase. As a result, SU(3) gauge symmetry is broken into the U(1) gauge symmetry
and the chiral symmetry is also broken.
Further increase of the chemical potential makes a quark pair
condensation $\epsilon^{ijk}\langle q_iq_j\rangle=(0,0,u)$ arise so that
the color superconducting phase ( 2CS ) appears. This state does not expel
the color magnetic field because the field does not couple with
the quark pair condensate.
In the phase of the coexistence of both states ( ferromagnetic
and superconducting states ), the chiral symmetry is restored, but
the U(1) gauge symmetry is broken due to the quark pair condensation.
This is a brief picture of the phase of the quark matter in the SU(3) gauge theory with 
two flavors. At much larger chemical potentials strange quarks
may play a role and they may form color flavor locking superconducting
phase with u and d quarks. In the phase 
the color ferromagnetic state disappears.

%Finally, we comment that although we obtain 
%the phase of the coexistence of 2CS and the color ferromagnetic state
%by taking $B\propto \lambda_3$ under
%a simple assumption of the small value of $g^2$, we may
%have such a phase
%with arbital values of $\theta$. This is because Abelian gauge fields
%associated with remaining $U(1)\times U(1)$ gauge symmetry may make
%quark pairs condense with the attracive force between them. 
%As a result, $U(1)\times U(1)$ gauge symmetry is broken into $U(1)$ gauge
%symmetry associated with the color magnetic field. So the coexistence
%of both phases
%may be possible.
%This possibility is under investigation.  
%We also mention that almost of all the above results do not depend on
%the quark mass. Especially, the conclusion of the existence of the ferromagnetic state 
%and the quantum Hall state of gluons holds irrespective 
%the quark mass. The only effect of the quark mass is 
%on the numerical details of our results such as
%the minimum chemical potential for the realization
%of the quantum Hall state, a critical chemical potential
%between the ferromagnetic phase and the coexistence phase of
%two states, e.t.c..

\hspace*{1cm}

We would like to express thanks to
Profs. T.~Kunihiro, T.~Hatsuda, H.~Suganuma and M.~Asakawa
for useful discussions. 
A.~I.~and M.~O.~express thanks to the member of theory 
group in KEK for their hospitality.
This work was partially supported by Grants-in-Aid of the Japanese Ministry
of Education, Science, Sports, Culture and Technology (No.06572).
T.~N. would like to thank the Japan Society of Promotion of Science (JSPS)
for financial support.


\begin{thebibliography}{99}
\bibitem{colors}K. Rajagopal and F. Wilczek, {\tt hep-ph/0011333}.
\bibitem{colorf}A. Iwazaki and O. Morimatsu, {\tt nucl-th/0304005} to be published in
  Phys. Lett. B
%\bibitem{monopole}S. Mandelstam, Phys. Lett. 53B 476 (1975).\\
%G. tHooft, Nucl. Phys. B190 455 (1981).
\bibitem{savidy}G.K. Savvidy, Phys. Lett. {\bf 71B}, 133 (1977).\\
H. Pagels, Lecture at Coral Gables, Florida, 1978.
\bibitem{nielsen}N.K. Nielsen and
P. Olesen, Nucl. Phys. {\bf B144}, 376 (1978);
Phys. Lett. {\bf 79B}, 304 (1978).
\bibitem{qh}The Quantum Hall Effect, 2nd Ed., edited by R.E. Prange
  and S.M. Girvan ( Springer-Verlag, New York, 1990 ).\\
Quantum Hall Effects, edited by Z.F. Ezawa ( World Scientific ).\\
Z.F. Ezawa, M. Hotta and A. Iwazaki, Phys. Rev. {\bf B46},  7765 (1992);
Z.F. Ezawa and A. Iwazaki, J. Phys. Soc. Jpn {\bf 61}, 4133 (1990). 
\bibitem{nn}J. Ambijorn, N.K. Nielsen and P. Olesen, Nucl. Phys. {\bf B152},
  75 (1979).\\
H.B. Nielsen and M. Ninomiya, Nucl. Phys. {\bf B156}, 1
  (1979).\\
H. B. Nielsen and P. Olesen, Nucl. Phys. {\bf B160}, 330 (1979).
\bibitem{abelian}Z.F. Ezawa and A. Iwazaki, Phys. Rev. {\bf D25}, 2681
  (1982).\\
T. Suzuki and I. Yotsuyanagi, Phys. Rev. {\bf D42}, 4257 (1990).
\bibitem{naka} private communication with Dr. T. Nakajima. 
\bibitem{spa}H.B. Nielsen and P. Olesen, Nucl. Phys. {\bf B160}, 380 (1979).
\bibitem{seme}G.W. Semenoff and P. Sodano, Nucl. Phys. {\bf B328}, 753 (1989).
%\bibitem{wil}M. Alford, K. Rajagopal and F. Wilczek, Phys. Lett. B422
%  247 (1998).
\bibitem{zhang}S.C. Zhang, H. Hansson and S. Kivelson, 
Phys. Rev. Lett. {\bf 62}, 82 (1989).
\bibitem{ebert}D. Ebert, V.V. Khudyakov, V.Ch. Zhukovsky and
  K.G. Klimenko, Phys. Rev. {\bf D65}, 054024 (2002).
\end{thebibliography}
\end{document}